\documentclass[aps,twocolumn,superscriptaddress,showpacs,pre,floatfix]{revtex4-1}
\usepackage{graphicx,color,mathtools}

\definecolor{verdon}{cmyk}{1,0.5,1,0}
\definecolor{blue}{cmyk}{0.8,0.8,0,0.}
\definecolor{red}{cmyk}{0.2,1,1,0.0}

\def\lapprox{\mathrel{\mathop  {\hbox{\lower0.5ex\hbox{$\sim$}
\kern-1.1em\lower-0.7ex\hbox{$<$}}}}}
\def\gapprox{\mathrel{\mathop  {\hbox{\lower0.5ex\hbox{$\sim$}
\kern-1.1em\lower-0.7ex\hbox{$>$}}}}}

\begin{document}

\title{\color{verdon} Expectations for high energy diffuse galactic neutrinos for different cosmic ray distributions}

\author{Giulia Pagliaroli}
\author{Carmelo Evoli}
\affiliation{Gran Sasso Science Institute, 67100 L'Aquila (AQ), Italy}
\author{Francesco Lorenzo Villante}
\affiliation{L'Aquila University, Physics and Chemistry Department, 67100 L'Aquila, Italy}
\affiliation{INFN, Laboratori Nazionali del Gran Sasso, 67100 Assergi (AQ),  Italy}

\begin{abstract}
The interaction of cosmic rays with the gas contained in our Galaxy is a guaranteed source of diffuse high energy neutrinos.
We provide expectations for this component by considering different assumptions for the cosmic ray distribution in the
Galaxy which are intended to cover the large uncertainty in cosmic ray propagation models.  
We calculate the angular dependence of the diffuse galactic neutrino flux and the corresponding rate of High Energy Starting
Events in IceCube by including the effect of detector angular resolution.
Moreover we discuss the possibility to discriminate the galactic component from an isotropic astrophysical flux. 
We show that a statistically significant excess of events from the galactic plane in present IceCube data would favour models in which
the cosmic ray density in the inner galactic region is much larger than its local value, thus bringing relevant information
on the cosmic ray radial distribution.
\end{abstract}

\maketitle

\section{Introduction}
\label{Introduction}

In four years of data taking, the IceCube detector has detected 54 High
Energy Starting Events (HESE) with deposited energy between 20 TeV and
2 PeV  which are compatible with an astrophysical population of high energy
neutrinos \cite{Aartsen:2013jdh,Aartsen:2014gkd,Aartsen:2015ivb,
Aartsen:2015rwa, Aartsen:2015knd, Botner}. The observed 
excess has been also confirmed by independent observation of 
upward going passing muons in IceCube \cite{Aartsen:2015rwa}.
The origin of these neutrinos is still unknown and potential sources 
include supernova remnants \cite{Kistler:2006h}, pulsars \cite{burgio}, 
active galactic nuclei \cite{Murase} and starburst galaxies \cite{Lacki}.      
Dedicated searches for point-like or extended sources have been performed
by IceCube \cite{ultimo,Aartsen:2014cva}; however, at present, 
no significant clustering or correlation of event arrival directions 
with potential source distributions 
has been found, thus leaving open the possibility of a diffuse
astrophysical neutrino population.

The isotropic distribution of the IceCube high energy events can be considered as an argument 
in favor of extragalactic origin of the signal \cite{waxman,tamborra,murasenew}. 
Recent works, however, pointed out that IceCube
data do not exclude (or are even better fitted) by allowing for a non negligible contribution 
of galactic origin \cite{Palladino, Ahlers, Neronov, Neronov2,Troitsky}.
It is known that the interactions of Cosmic Rays (CR) 
with the interstellar medium is a guaranteed source of a diffuse
neutrinos in our Galaxy.  
The calculation of this component is, however, quite uncertain because it
requires the knowledge  of the CR distribution in all the regions of the Galaxy 
where the gas density is not negligible.   
The standard approach relies on local measurements 
and on the solution of CR transport equations by assuming constant 
diffusion in the whole Galaxy \cite{berezinskii}.
The recent results provided by Fermi-LAT \cite{Acero} may challenge 
this scenario since they seems to indicate a dependence of the CR spectrum 
and distribution on the distance from the Galactic Center, as it is e.g.
expected in CR propagation model characterised 
by radially dependent transport properties {\cite{Evoli,Grasso,Morlino}}.

In this paper, we describe a self-contained
calculation of the diffuse galactic neutrino flux that allow us
to discuss the expectations, uncertainties and
detectability of this component in general terms, without entering  
in the complex problem of CR propagation in the Galaxy. 
We calculate the angular dependence of the galactic neutrino flux and 
the corresponding rate of High Energy Starting Events (HESE) in
IceCube by considering different assumptions for the CR density in the
Galaxy that are intended to cover the large uncertainty in CR
propagation models. Namely, we assume that CR distribution is
homogenous in the Galaxy ({\em Case A}), that it follows the 
distribution of galactic CR sources ({\em Case  B}) and that 
it has a spectral index that depends on the galactocentric distance 
({\em Case C}).
 We then discuss the perspectives for the extraction of a 
 galactic neutrino signal from IceCube HESE data showing that 
 only {\em Case C} gives a non negligible chance of detection.

The plan of the paper is the following. In the next section, we
introduce the main ingredients of our calculation. 
In sect.~\ref{CRflux}, we discuss our assumptions for the CR distribution
in  the Galaxy. 
In sect.~\ref{neutrinoflux}, we calculate the neutrino flux as a function of
neutrino energy and arrival direction.
In sect.\ref{EventRates}, we calculate the expected rates of HESE in
IceCube, taking into account the different angular resolution for
shower and track events, and we discuss the perspectives for the 
extraction of the galactic neutrino signal.
In sect.~\ref{Conclusions}, we summarise our results.

\section{Notations}
\label{Notations}

The flux of high energy neutrinos produced at Earth by interactions of CRs 
with the gas contained in the galactic disk can be written as:

\begin{eqnarray}
\nonumber
& &  \varphi_\nu (E_\nu,\hat{n}_\nu) = \frac{1}{3}
\sum_{\ell=e,\mu,\tau}\left[\int_{E_\nu}^{\infty} d E \, \right.
\frac{d\sigma_\ell(E ,E_{\nu})}{dE_{\nu}}\, \\
& & \left. 
\int_{0}^{\infty}  dl \,\varphi_{\rm CR} (E,{\bf r}_{\odot}+ l \, \hat{n}_\nu )\,
n_{\rm H} ({\bf r}_{\odot}+l \, \hat{n}_\nu )\right]
\label{nuflux}
\end{eqnarray}
where $E_\nu$ and $\hat{n}_\nu$ indicate the neutrino energy and arrival
direction, $d\sigma_\ell/dE_\nu$ is the differential cross section for
production of neutrinos $\nu_\ell$ and antineutrinos
$\overline{\nu}_\ell$ by a nucleon of energy $E$ 
in nucleon-nucleon collision. The function $\varphi_{\rm CR} (E,{\bf r})$ represents
the differential CR flux (see next section), $n_{\rm H}({\bf r})$ is the gas
density distribution and $ r_{\odot}=8.5$ kpc 
is the position of the Sun. 
In the above relation, we assumed that, due to neutrino mixing, 
the neutrino flux at Earth is equally distributed among the different
flavours. This approximation is valid with few $\%$ accuracy, as it is
discussed in e.g.~\cite{nuflavor} and it is completely adequate for our purposes.  
For the nucleon-nucleon cross section, following
\cite{kelner}, we assume:
\begin{equation}
\sum_{\ell=e,\mu,\tau} \frac{d\sigma_\ell(E,E_{\nu})}{dE_{\nu}} = \frac{\sigma(E)}{E}\; 
F\left(x,E \right),
\end{equation}
where $x=E_\nu/E $ and the total inelastic cross section $\sigma(E)$ is given by:
\begin{equation}
\sigma(E) = 34.3 +1.88\ln(E/{\rm 1 TeV}) + 0.25 \ln(E/{\rm 1 TeV})^2
\;\; {\rm mb.}
\nonumber
\end{equation}
The adimensional distribution function $F(x,E)$ is given by:
\begin{equation}
F\left(x,E \right) =  \left[F_{\nu_\mu} 
\left(x,E  \right) + F_{\nu_e} \left(x,E  \right) \right],
\end{equation}
where $F_{\nu_\mu} \left(x,E  \right)$ and $F_{\nu_e}  \left(x,E  \right)$
are described (with 20\% accuracy) by the analytic formulas given in \cite{kelner}.

The galactic distribution of the gas density, $n_{\rm H}({\bf r})$, 
is taken from the public GALPROP code~\cite{Galprop} and described in~\cite{SM}.
This is given as a sum of atomic, $H_I$, and molecular $H_2$ hydrogen. 
For the latter we adopt a conversion factor with respect to the $CO$ density as given 
by~\cite{Yang} from a fit of the diffuse $\gamma$ emission.
Finally, we assume that Helium contributes to the Galactic gas with a constant density ratio of $0.11$ with respect to total Hydrogen.

\section{The CR flux}
\label{CRflux}

In order to predict the neutrino flux at $E_\nu\simeq 100 \,{\rm TeV}$, we
need to know the CR flux at $E \simeq 20 \, E_\nu = 2 \,{\rm  PeV}$.  
At the Sun position, the CR flux is constrained by observational data and 
we can write: 
\begin{equation}
\nonumber
\varphi_{\rm CR,\odot} (E) \equiv \sum_A A^2
\frac{d\phi_A}{dE_A d\Omega_A}(A E)
\end{equation}
where $d\phi_A / dE_A d\Omega_A$ is the differential flux at Earth 
of a given nuclear species, $A$ represents the nuclear mass number and we
considered that the energy of the nucleus is $E_A = A \, E$. 
We use the parameterisations for $d\phi_A / dE_A d\Omega_A$ given by 
\cite{Ahlers} that are obtained by fitting the CREAM \cite{CREAM},
KASCADE \cite{KASCADE} and KASCADE-Grande \cite{KASCADEGrande} data
in the energy range $E_A\sim 1 -  10^6  \,{\rm TeV}$ and assuming that the 
dominant contributions to the nucleon flux $\varphi_{\rm CR,\odot} (E)$
are provided by ${\rm H}$ and $^4{\rm He}$
nuclei. 
Note that, if large fluxes of heavy nuclei are introduced at expenses 
of ${\rm H}$ and $^4{\rm He}$ components (i.e. maintaining $\sum_A d\phi_A / dE_A d\Omega_A = {\rm
  const}$), the nucleon flux $\varphi_{\rm CR,\odot} (E)$ is reduced 
because the CR spectral distributions are decreasing with energy
faster than $E^{-2}$.

The local determination $\varphi_{\rm CR,\odot} (E) $ has to be
related to the CR flux in all the regions of the Galaxy where the
gas density is not negligible. We consider here three different
prescriptions of increasing complexity that correspond to 
different amounts of energy stored in CR. 

\vspace{0.5 cm}

\noindent
{\em Case A:} We assume that the CR flux is homogenous in the Galaxy,
i.e. we write:
\begin{equation}
\varphi_{\rm CR} (E,{\bf r}) \equiv \varphi_{\rm CR,\odot} (E) 
\end{equation}
In this assumption, the neutrino flux can be expressed as:
\begin{equation}
\varphi_\nu (E_\nu,\hat{n}_\nu) = {\mathcal F}_\nu (E_\nu) \,
{\mathcal A}(\hat{n}_\nu)
\label{nufluxFact}
\end{equation}
where the function that contains the angular dependence:
\begin{equation}
{\mathcal A} (\hat{n}_\nu) = \frac{1}{{4\pi \mathcal N}_{\rm H} } \, \int_{0}^{\infty}  dl \, n_{\rm H} ({\bf r}_{\odot}+l \, \hat{n}_\nu )
\end{equation} 
is proportional to the column density of the gas along a given
direction, the normalisation parameter ${\mathcal N}_{\rm H}$ 
is the average column density of the gas given by:
\begin{equation}
{\mathcal N}_{\rm H} = 
\frac{1}{4\pi}  \int d^3r\; 
\frac{n_{\rm H} ({\bf r}_{\odot}+ {\bf r})}{ r^2} = 
2.19 \times 10^{21} \;{\rm cm}^{-2}
\end{equation}
and the function:
\begin{equation}
{\mathcal F}_\nu (E_\nu) = \frac{4 \pi}{3} {\mathcal N}_{\rm H}
\int_{E_\nu}^{\infty} \frac{d E}{E} \,
\sigma(E) \, F\left(\frac{E_\nu}{E },E \right) 
 \varphi_{\rm CR,\odot} (E)
\end{equation}
is the neutrino flux integrated over arrival directions.

\vspace{0.5 cm}

\begin{figure}
\begin{center}
\includegraphics[width=.46\textwidth,angle=0]{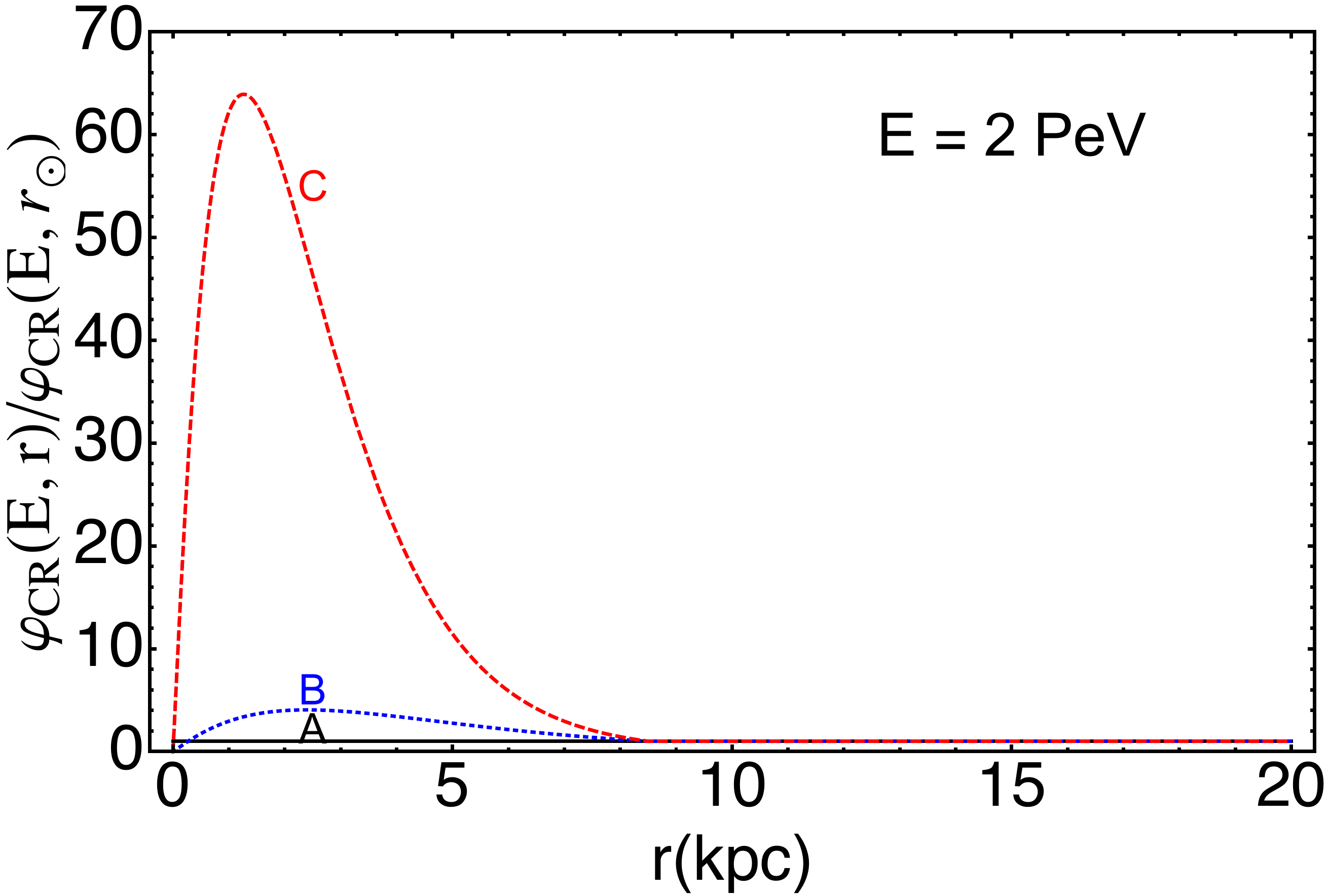}
\end{center}
\caption{\small\em The CR flux at $E = 2\,{\rm PeV}$ as a function of the
distance from the galactic center in the three considered
scenarios. We assume that the CR flux is approximately constant along
the galactic latitudinal axis. See text for details.\label{fig1}}
\end{figure}

\noindent
{\em Case B:} We assume that the CR flux scales proportionally to
the distribution of CR sources, as it is roughly expected if CR escape
is much faster from the halo than radially.
%
Namely, we write:
\begin{equation}
\varphi_{\rm CR} (E,{\bf r}) \equiv \varphi_{\rm CR,\odot} (E) \,
g({\bf r})
\label{CaseB} 
\end{equation}
where:
\begin{equation}
g({\bf r}) = \frac{n_{\rm S}({\bf r})}{n_{\rm S}({\bf r}_{\odot})}
\label{gS}
\end{equation}
and $n_{\rm S}({\bf r})$ describes the CR source density.
In this assumption, the neutrino flux can still be factorised as in
eq.(\ref{nufluxFact}) but the function ${\mathcal A} (\hat{n}_\nu)$ is replaced by the function: 
\begin{equation}
\mathcal{B}(\hat{n}_\nu) = \frac{1}{{4\pi \mathcal N}_{\rm H}
} \int_{0}^{\infty}  dl \, n_{\rm H} ({\bf
  r}_{\odot}+l \, \hat{n}_\nu ) g ({\bf r}_{\odot}+l \, \hat{n}_\nu)
\end{equation}
We take the SNRs distribution parameterised by Green et
al.~\cite{SNRs} as representative for the source density $n_{\rm S}({\bf r})$. 
However, since it is known, e.g. from Fermi-LAT observations of the
galactic $\gamma$-ray emission \cite{Acero}, that in the outer region of the Galaxy
the CR density drops slower than what one would expects from SNRs \cite{SNRs} (or
pulsars \cite{Pulsars}) distributions, we use eqs.(\ref{CaseB}) only for
galactocentric distances $r\le r_\odot$.
Moreover, since the CR diffusion length is expected to be larger than
both the thickness of the Galactic Disk and of that of the SNRs distribution, 
we assume that the CR flux is constant along the galactic latitudinal 
axis. 
In these assumptions, the function $g({\bf r})$ is given in galactic
cylindrical coordinates by:
\begin{equation}
g(r, z)  = \left(\frac{r}{r_{\odot}}\right)^\gamma \exp\left(-\beta \frac{r-r_{\odot}}{r_{\odot}}\right)
\end{equation}   
where $\gamma=1.09$, $\beta =3.87$ and we neglected the dependence on $z$. 
The function $g({\bf r})$ is shown by the blue line in Fig.\ref{fig1}.
We see that the CR density is larger by a factor $\sim 4$ 
at distances $r=2-3$ kpc from the galactic center with respect to its
local value. To provide a quantitative comparison, we note that the
energy stored in CR contained at $r\le r_\odot$ is a factor $2.3$
larger than what obtained in the assumption of CR homogeneity.

\vspace{0.5cm}

\noindent
{\em Case C:}  We consider the possibility that the CR spectral
distribution depends on the position, as it has been recently observed by
Fermi-LAT at low energies~\cite{Gaggero,Yang,Acero}.
%
%
To this purpose, we write:
\begin{equation}
\varphi_{\rm CR} (E,{\bf r}) \equiv \varphi_{\rm CR,\odot} (E) \,
g({\bf r}) \, h(E,{\bf r}) 
\end{equation}
where the function:
\begin{equation}
h(E,{\bf r}) =\left( \frac{E}{\overline{E}} \right)^{\Delta({\bf r})}
\label{h}
\end{equation}
introduces a position-dependent variation $\Delta({\bf r})$ of the CR
spectral index. 
The pivot energy in eq.(\ref{h}) is taken as
$\overline{E}=20\;\rm{GeV}$,  
since it is observed \cite{Acero, Morlino}
that the integrated CR density above 20 GeV roughly follows 
the function $g({\bf r})$ defined in {\em Case B}. 

Having no direct informations on the radial distribution of 
high energy CR, we are forced to rely on extrapolations from low
energy data and even a relatively small $\Delta({\bf r})$ may
introduce a large error.  
At the energy $E_{\rm CR} = 2\,{\rm PeV}$ which is most
relevant for neutrino telescopes, the major effect of the function $h(E,{\bf r})$
is to rescale the CR flux by a factor
\begin{equation}
\overline{h}({\bf r}) =\left( \frac{E_{\rm CR}}{\overline{E}} \right)^{\Delta({\bf r})}
\end{equation}
that depends on the position but that can be approximately 
considered energy-independent. In this assumption, the neutrino flux is still given by 
eq.(\ref{nufluxFact}) but the function ${\mathcal A} (\hat{n}_\nu)$ is replaced by: 
\begin{equation}
\mathcal{C} (\hat{n}_\nu) =  
\frac{1}{{4\pi \mathcal N}_{\rm H}}
\int_{0}^{\infty}  dl \, n_{\rm H} ({\bf
  r}_{\odot}+l \, \hat{n}_\nu ) g ({\bf r}_{\odot}+l \, \hat{n}_\nu
)\, 
\overline{h}
({\bf r}_{\odot}+l \, \hat{n}_\nu).
\end{equation}

For our calculations, we take the function:
\begin{equation}
\Delta(r, z) = 0.3 \left(1-\frac{r}{r_{\odot}}\right)
\end{equation}
for $r\le r_{\odot}$, in galactic cylindrical coordinates, that 
is intended to reproduce the trend of the spectral index 
with $r$ observed by \cite{Acero} at 20 GeV
and that is also used by \cite{Grasso} in their 
phenomenological CR propagation model 
characterised by radially dependent transport properties. 
This corresponds to increasing the CR flux at $E_{\rm CR} =
2\, {\rm PeV}$ by a factor $(E_{\rm  CR}/{\overline{E}})^{0.3}\simeq
30$ close to the galactic center with respect to {\em Case B}. 

The product $g({\bf r})\overline{h}({\bf r})$ is shown by the red dashed line in Fig.~\ref{fig1} 
from which we see that the CR density at $2\,{\rm PeV}$ is larger by a factor $\sim 60$
at distances $r=2-3$ kpc from the galactic center with respect to its value at the Sun
position.  The energy stored in CR above 2 PeV at distances $r\le
r_\odot$  is a factor 14 larger than what obtained in the assumption of
CR homogeneity (i.e. {\em case A}).

\section{The neutrino flux}
\label{neutrinoflux}

In the three cases described above, the flux of high energy neutrinos
and antineutrinos of each flavour at Earth can be written as: 
\begin{equation}
\varphi_{\nu}(E_\nu,\hat{n}_\nu) = {\mathcal F}(E_{\nu})\, 
{\mathcal I}(\hat{n}_\nu)
\end{equation}
where ${\mathcal I} = {\mathcal A},\,{\mathcal B}, {\mathcal C}$
depending on the considered scenario. Being 
$\overline{{\mathcal A}} \equiv\int d\Omega\, {\mathcal A}
(\hat{n}) =1 $, the function  ${\mathcal
  F}(E_{\nu})$ represent the angle-integrated neutrino flux 
in the {\em Case A} (i.e. uniform CR density).
For neutrino energies $E_\nu = 10\,{\rm TeV} - 1\,{\rm PeV}$, 
this is well approximated by:
\begin{equation}
{\mathcal  F}(E_{\nu}) = f  \left[\frac{E_\nu}{100 \, {\rm TeV}}\right]^{-\alpha(E_\nu)}
\end{equation} 
where $f =4.76\times 10^{-7}\,{\rm GeV}^{-1}\,{\rm m}^{-2}\,{\rm
  y}^{-1}$ and the spectral index is given by:
\begin{equation}
\alpha(E_\nu) = 2.65 + 0.13 \log_{10}\left( E_\nu / 100 \, {\rm TeV} \right).
\end{equation}
Note that the functions ${\mathcal B}(\hat{n}_\nu)$ and ${\mathcal C}(\hat{n}_\nu)$ are 
not normalized.  The integrated neutrino fluxes in these scenarios are thus given by:
\begin{equation}
\phi_\nu(E_\nu)=\overline{{\mathcal I}} \, {\mathcal F}_\nu (E_\nu)
\end{equation}
where the factors $\overline{{\mathcal I}} \equiv\int d\Omega\, {\mathcal I}
(\hat{n}) $ are equal to $\overline{{\mathcal B}}=1.23$ and
$\overline{{\mathcal C}}=2.34$, respectively. 

The angle-integrated fluxes can be compared with the isotropic flux: 
\begin{equation}
{\mathcal F}_{\rm iso} (E_\nu) = f_{\rm iso} \, \left[\frac{E_\nu}{100 \,
    {\rm TeV}}\right]^{-2.58}  
\end{equation}
where $f_{\rm iso} = 8.72\times 10^{-6}\,{\rm GeV}^{-1}\,{\rm
  m}^{-2}\,{\rm  y}^{-1}$, that corresponds to the HESE
event rate observed by IceCube in four years data taking
\cite{Icecube4}. 
At the neutrino energy  $E_{\nu}=100 \,{\rm TeV}$
that provides the most relevant contribution to the HESE data
sample, the diffuse galactic neutrino component is equal to $5\%$, $7\%$ and 
$13\%$ in {\em Case A, B} and {\em C} respectively, of the isotropic
flux required to explain the 54 events observed by IceCube.
This component is thus not negligible but always subdominant and  
well consistent with the upper limit derived from \cite{Ahlers} by
fitting the event arrival directions.

\begin{figure*}
\begin{center}
\includegraphics[width=.47\textwidth,angle=0]{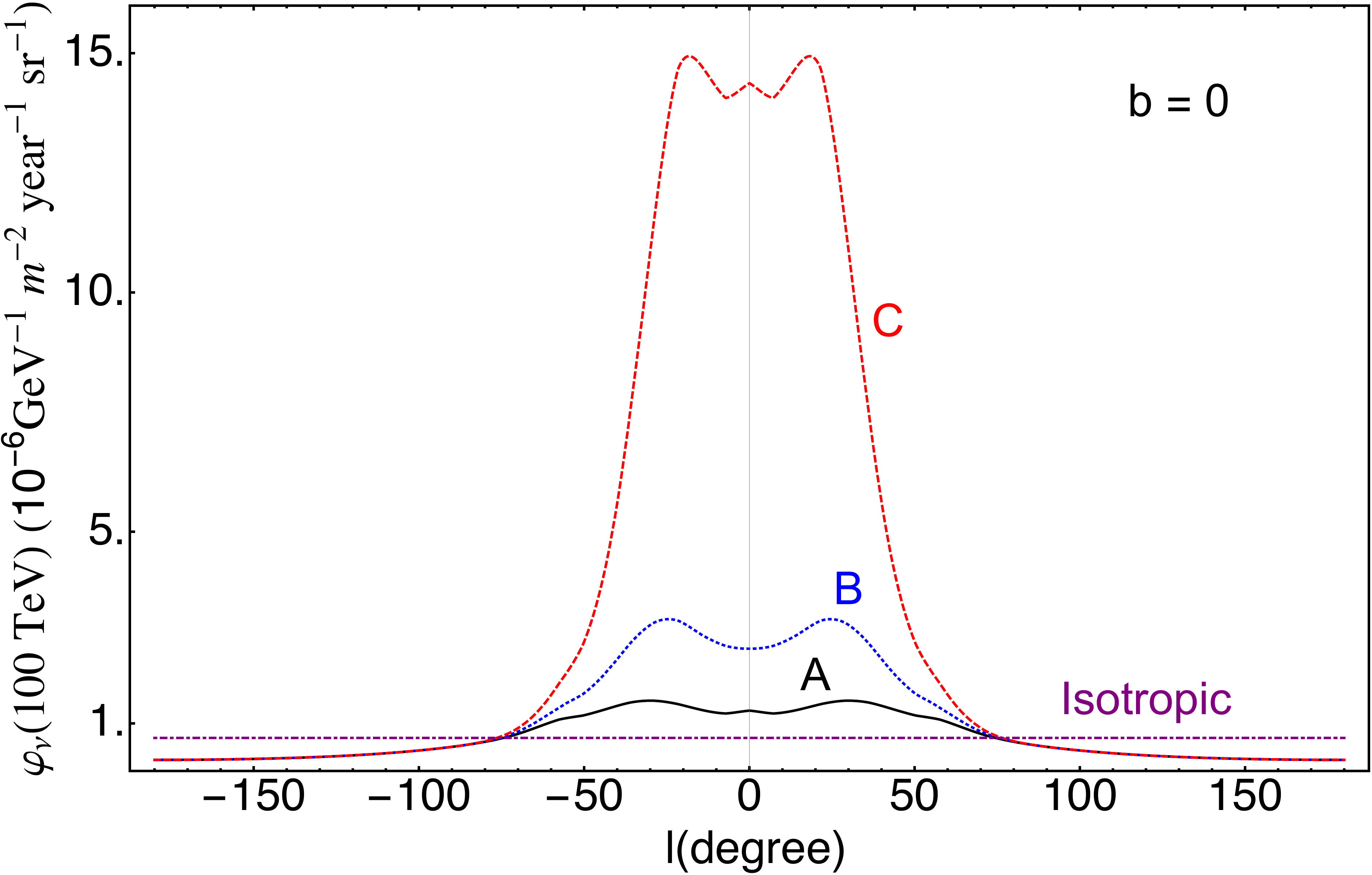}
\hspace{\stretch{1}}
\includegraphics[width=.47\textwidth,angle=0]{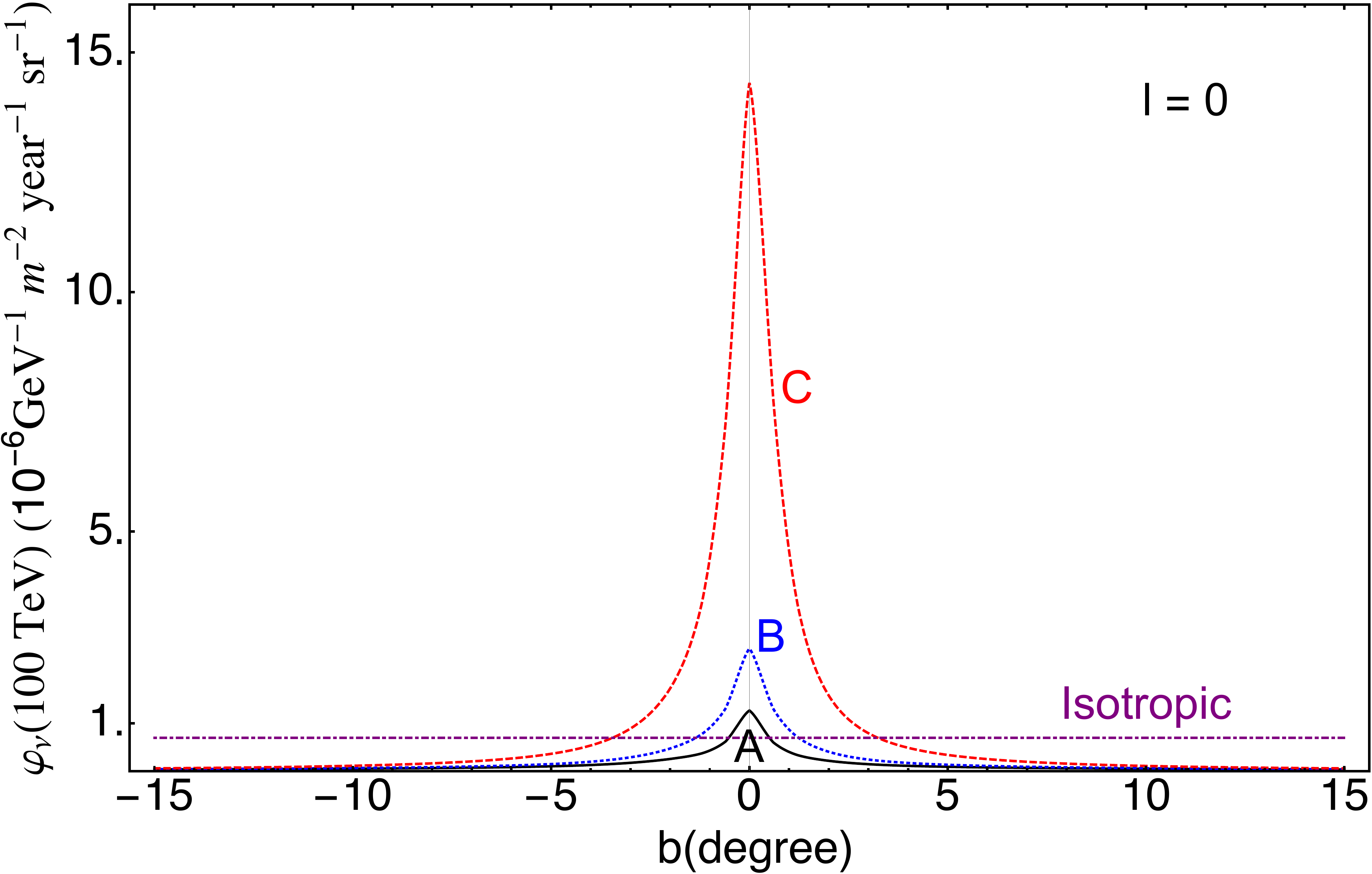}
\end{center}
\caption{The neutrino flux at $E_\nu=100 \,{\rm TeV}$ as a function of the Galactic longitude (left panel) and latitude (right panel) 
for the three different models considered for CR distribution. 
The solid black line corresponds to {\em Case A},
the blue dotted line corresponds to {\em Case B} 
and the red dashed line corresponds to {\em Case C}. 
The isotropic flux that reproduces IceCube HESE data is also 
reported for comparison with a purple dot-dashed line. \label{fig2}}
\end{figure*}

 The angular dependence of the flux in the three considered 
scenarios is shown in Fig.~\ref{fig2} as a function of the galactic 
longitude $l$ (left panel) and latitude $b$ (right panel).  
 We note that:\\
{\em i)} In the considered scenarios, it always exists a region, that
contains the galactic center, where the neutrino flux produced by CR
interacting with the gas contained in the galactic disk, is 
comparable or larger than the
isotropic contribution. Thus, the diffuse galactic neutrino component 
is, in principle, sufficiently intense to be detected. We recall that this
component is guaranteed by the existence of CR at PeV energies, as
it is observed e.g. by CREAM, KASCADE and KASCADE-Grande experiments.
Our calculations are based on the local determination of the CR flux
$\varphi_{\rm CR,\odot} (E)$ described in sect.\ref{CRflux}. We warn the reader
that other interpretations of the experimental data are possible
\cite{Gaisser}  which may decrease the neutrinos flux by a factor
$\sim 2$ \cite{Ahlers} without altering, however, our conclusions;\\
{\em ii)} The region where the diffuse galactic neutrino
component dominates is quite narrow. Even in the most optimistic {\em
  Case C}, the region where
$\varphi_{\nu}(E_\nu,\hat{n}_\nu) \ge {\mathcal F}_{\rm iso} (E_\nu) /
4\pi$ corresponds to $ | l | \le  70^\circ $ and $  |b| \le 3^\circ$. 
Thus, the optimal detector should have a good pointing capability
in order to avoid diluting the signal below the isotropic background.
Unfortunately, the IceCube HESE data set is dominated by showers events
that do not allow to reconstruct the neutrino arrival direction with 
sufficient accuracy (see next section, for a detailed discussion).\\
{\em iii)}  The angular distributions are quite different in the three
considered cases. The maximal emission is always achieved 
for $l \simeq \pm 25^\circ$ and $b = 0^\circ$ but the neutrino fluxes
may differ by large factors for $|l| \le 90^\circ$. 
To be quantitative, the flux from the galactic
center is larger by a factor $\sim 2$ and $\sim11$ in {\em Case B} and {\em C}
respectively, with respect to the value obtained in the assumption of
uniform CR density (i.e. {\em Case A}). In perspective, this
could provide an handle to discriminate among different 
scenarios, in an ideal detector with sufficient statistics and 
good pointing capability.

\section{Event rate in IceCube}
\label{EventRates}

The number of HESE event expected in IceCube can be calculated by using the
effective areas $A_\ell \left(E_\nu,\hat{n}_\nu\right)$ provided by \cite{IceCubeAeff} according to:
\begin{eqnarray}
\nonumber
N_{\rm S} &=&  T \int dE_\nu \int d\Omega_\nu \;
\varphi_{\nu}(E_\nu,\hat{n}_\nu)   \left[
A_{e}\left(E_\nu,\hat{n}_\nu\right) + \right.\\
& & \hspace{-0.5cm}\left. A_{\mu}\left(E_\nu,\hat{n}_\nu\right) (1-\eta) +
A_{\tau}\left(E_\nu,\hat{n}_\nu \right)  \right]\\
\nonumber
 N_{\rm T} &=&  \eta \, T \int dE_\nu \int d\Omega_\nu \;
\varphi_{\nu}(E_\nu,\hat{n}_\nu)  A_{\mu}\left(E_\nu,\hat{n}_\nu\right) 
\end{eqnarray} 
where $T$ is the observation time and we estimated the separate
contributions of shower ($N_{\rm S}$) and track ($N_{\rm T}$) events. 
In the above relation, we assume that neutrinos and antineutrinos 
are equally distributed among the different flavours, as it is
expected due to flavour oscillations; moreover,  we indicate with
$\eta \simeq 0.8$ the probability that a muon neutrino interacting 
in IceCube produces a track event, as it was estimated in \cite{nuflavor}.
This parameter is somewhat uncertain and may be reduced by systematic 
tracks misidentification error \cite{mis}. We remark,
however, that the specific value of $\eta$ does not alter our main conclusions.

\begin{table}
\centering
\caption{The track and shower HESE rates expected in IceCube for the three 
different models considered in the text and for the isotropic flux
observed by IceCube.  We also show the separate contributions from Northern and Southern hemisphere.   
 \label{tab1}}
\begin{tabular}{lcccc}
\hline 

 & \multicolumn{4}{c}{$N/T - {\rm counts \cdot y^{-1}}$} \\ 
\cline{2-5}
& {\em Showers} & {\em Tracks} & {\em North} & {\em South} \\ \hline
{\em Case A} &  0.40 & 0.07 & 0.18 & 0.29 \vspace{0.5mm}\\
{\em Case B}  &  0.50 & 0.09 & 0.20 & 0.39 \vspace{0.5mm}\\
{\em Case C}  &  1.01 & 0.19 & 0.27 & 0.92 \vspace{0.5mm}\\
\hline
{\em Isotropic} \hspace{0.5mm} & 8.33 & 1.61 & 4.13 & 5.80 \vspace{0.5mm}\\
\end{tabular}
\end{table}

The event rates corresponding to the three scenarios considered in
this paper are given in Tab.\ref{tab1} where we also give the separate 
contributions from the Northern and Southern hemisphere, calculated by
taking into account the angular resolution of the IceCube detector 
as it is described below.  In the assumption of uniform CR density ({\em
  Case A}), one obtains a total event rate $(N_{\rm S}+N_{\rm T})/T =
0.47 \,{\rm y}^{-1}$. 
For {\em  Case B} and {\em  Case C}, the predicted event rates are $(N_{\rm S}+N_{\rm T})/T = 0.60 \,{\rm
  y}^{-1}$  and $(N_{\rm S}+N_{\rm T})/T= 1.2\,{\rm y}^{-1}$, respectively.
For comparison, the isotropic flux corresponds to an integrated rate
$(N_{\rm S}+N_{\rm T})/T = 9.9 \,{\rm y}^{-1} $. 
We also see that the North-South asymmetry depends on the considered
scenario, being maximal and equal to $\sim55\%$ for {\em Case C}, 
as a result of a more pronounced emission from the inner Galactic
region.  In view of the smallness of the diffuse galactic neutrino
contribution, 
it appears however unplausible that this component may
introduce of a large  North-South asymmetry 
in the complete  IceCube HESE data sample (see \cite{Palladino} for a discussion). 

The angular distribution of events can be estimated by:
\begin{eqnarray}
\nonumber
\frac{dN_{\rm S}(\hat{n})} {d\Omega} &=&  T 
\int dE_\nu 
\int d\Omega_\nu\;
G_{\rm S}(\hat{n},\hat{n}_\nu)
\varphi_{\nu}(E_\nu,\hat{n}_\nu)  \\
\nonumber
& & \hspace{-1.cm}
\times
\left[
A_{e}\left(E_\nu,\hat{n}_\nu\right) +
A_{\mu}\left(E_\nu,\hat{n}_\nu\right) (1-\eta) +
A_{\tau}\left(E_\nu,\hat{n}_\nu \right)  \right]\\
\nonumber
\frac{dN_{\rm T}(\hat{n})} {d\Omega} &=&  \eta \; T 
\int dE_\nu  \int d\Omega_\nu \, G_{\rm T}(\hat{n},\hat{n}_\nu)\\
& & \varphi_{\nu}(E_\nu,\hat{n}_\nu) A_{\mu}\left(E_\nu,\hat{n}_\nu\right) \\
\nonumber
\end{eqnarray}
where the function $G_{\rm S}(\hat{n},\hat{n}_\nu)$ ($G_{\rm T}(\hat{n},\hat{n}_\nu)$)
describes the angular resolution, i.e. the probability that a shower
(track) event with a {\em reconstructed} 
direction $\hat{n}$ is produced by a a neutrino arriving from the direction $\hat{n}_{\nu}$. 
In principle, the angular resolution depends on the neutrino
energy, flavour, direction, etc.
Here, to avoid unnecessary complications, we take constant angular 
resolution for showers and track events, modeled as \cite{PRD}:
\begin{equation}
G_{\rm I}(\hat{n},\hat{n}_\nu) = \frac{m}{2\pi \delta n_{\rm I}^2}\exp
\left(-\frac{1-c}{\delta n_{\rm I}^2}\right)
\end{equation}
where ${\rm I=S,\, T}$, the parameter $m$ is a normalization factor,
$c \equiv \cos \theta = \hat{n} \,\hat{n}_\nu$ and the widths $\delta
n_{\rm S}$ and $\delta n_{\rm T}$ are calculated by requiring that $\theta\le
15^\circ$ at $68.3\%$ C.L. for showers and $\theta\le
1^\circ$ at $68.3\%$ C.L. for tracks \cite{Icecube4}.

\begin{figure*}
\begin{center}
\includegraphics[width=.47\textwidth,angle=0]{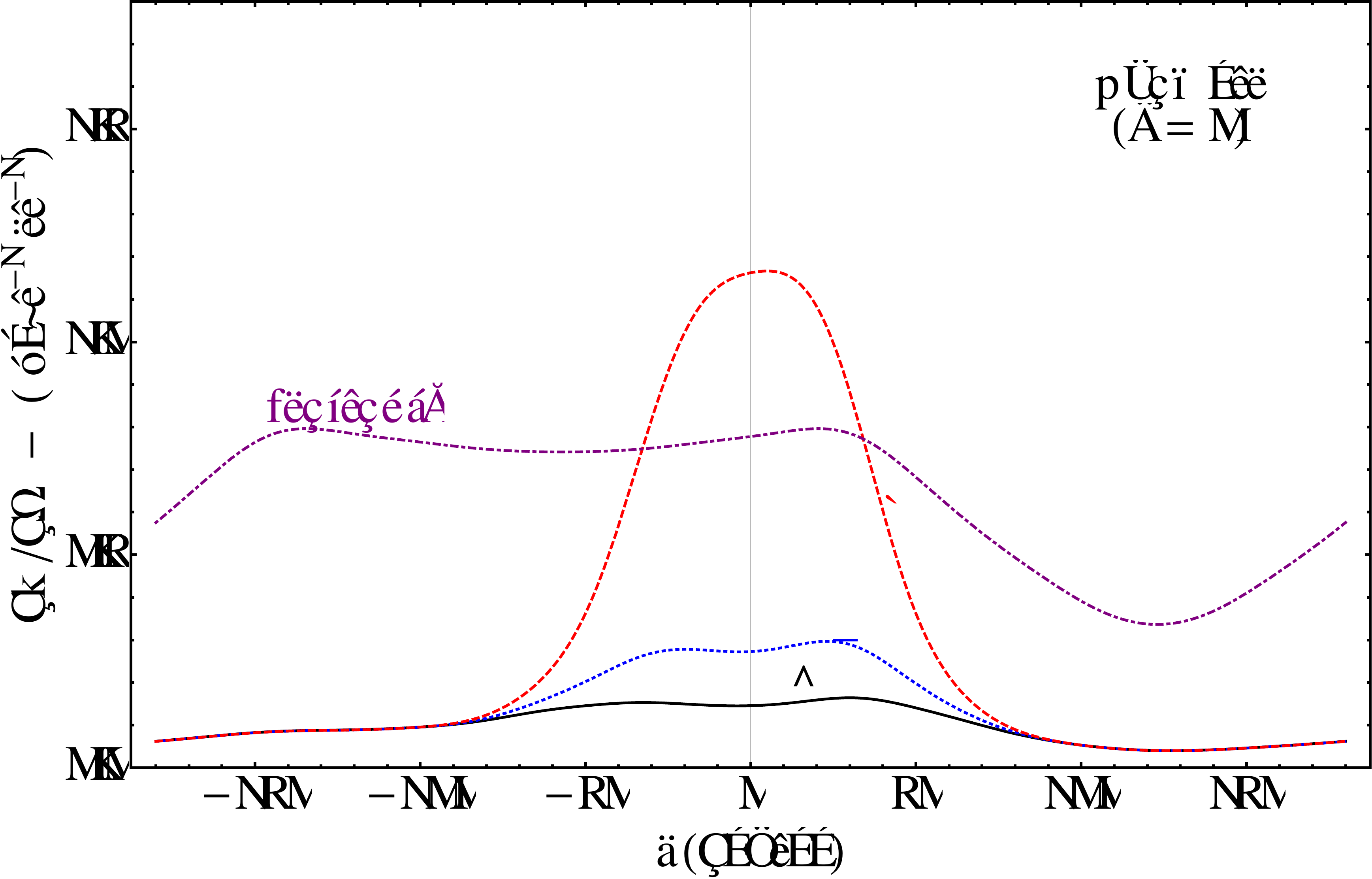}
\hspace{\stretch{1}}
\includegraphics[width=.47\textwidth,angle=0]{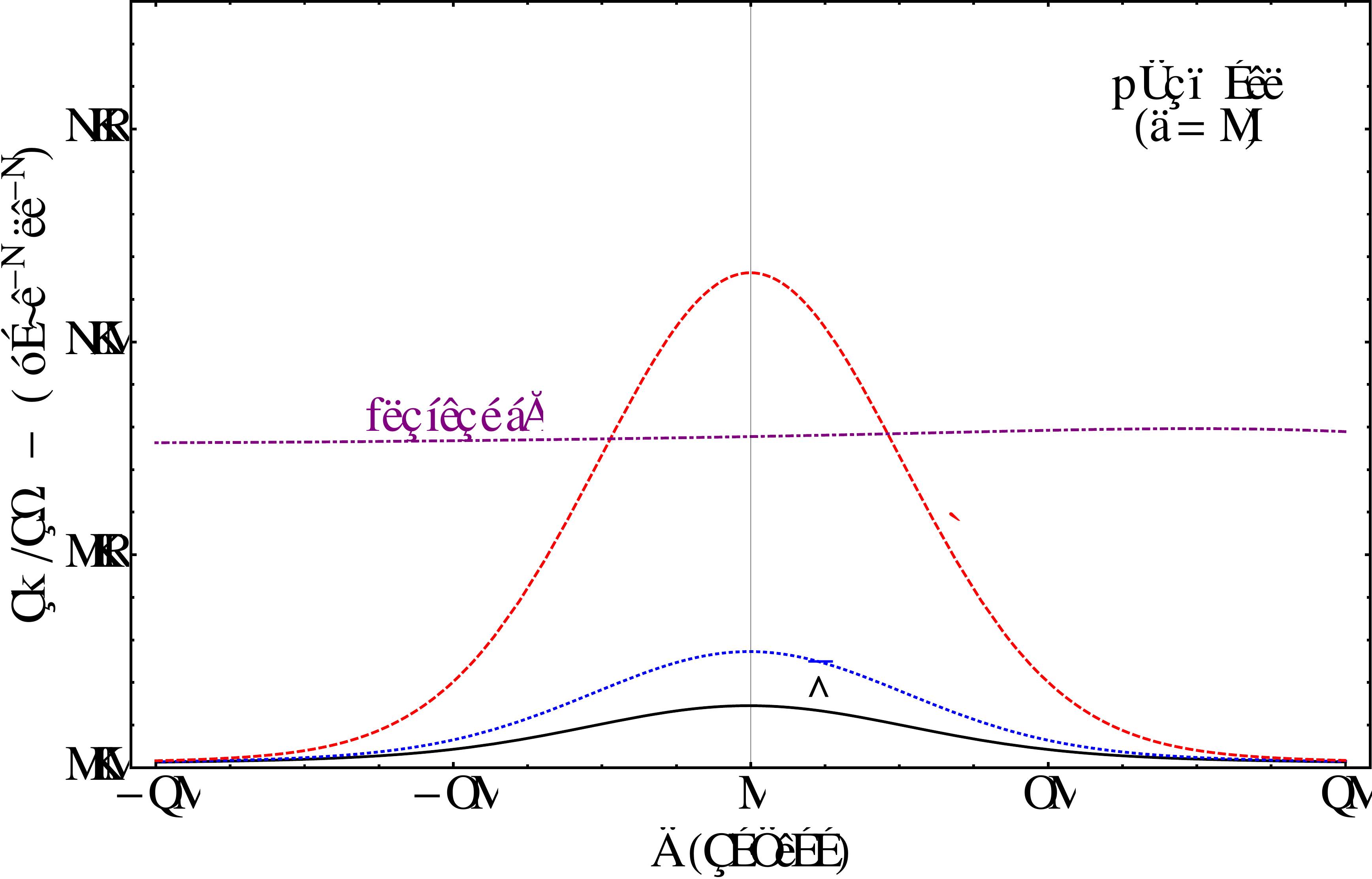} \\
\includegraphics[width=.47\textwidth,angle=0]{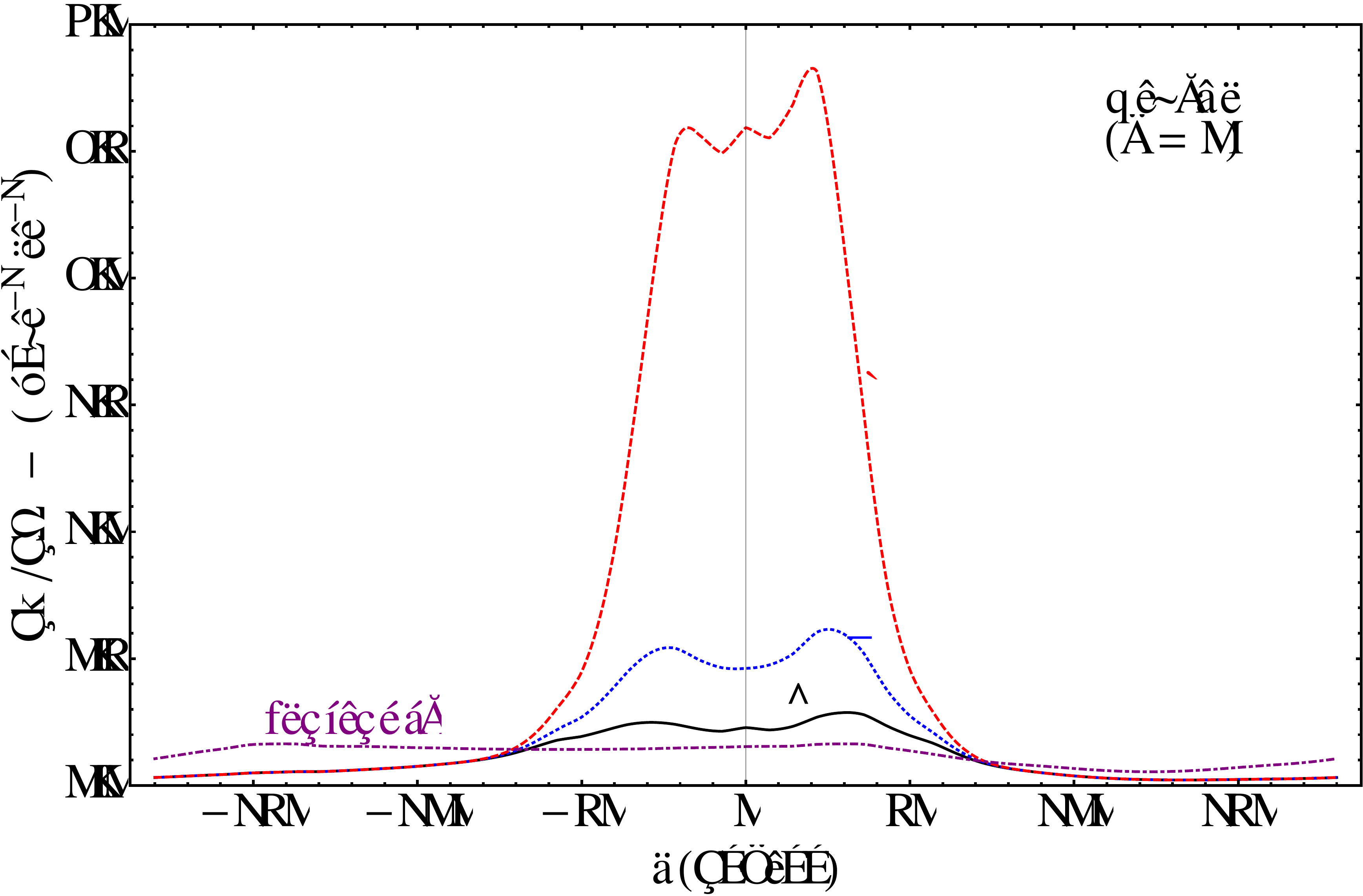}
\hspace{\stretch{1}}
\includegraphics[width=.47\textwidth,angle=0]{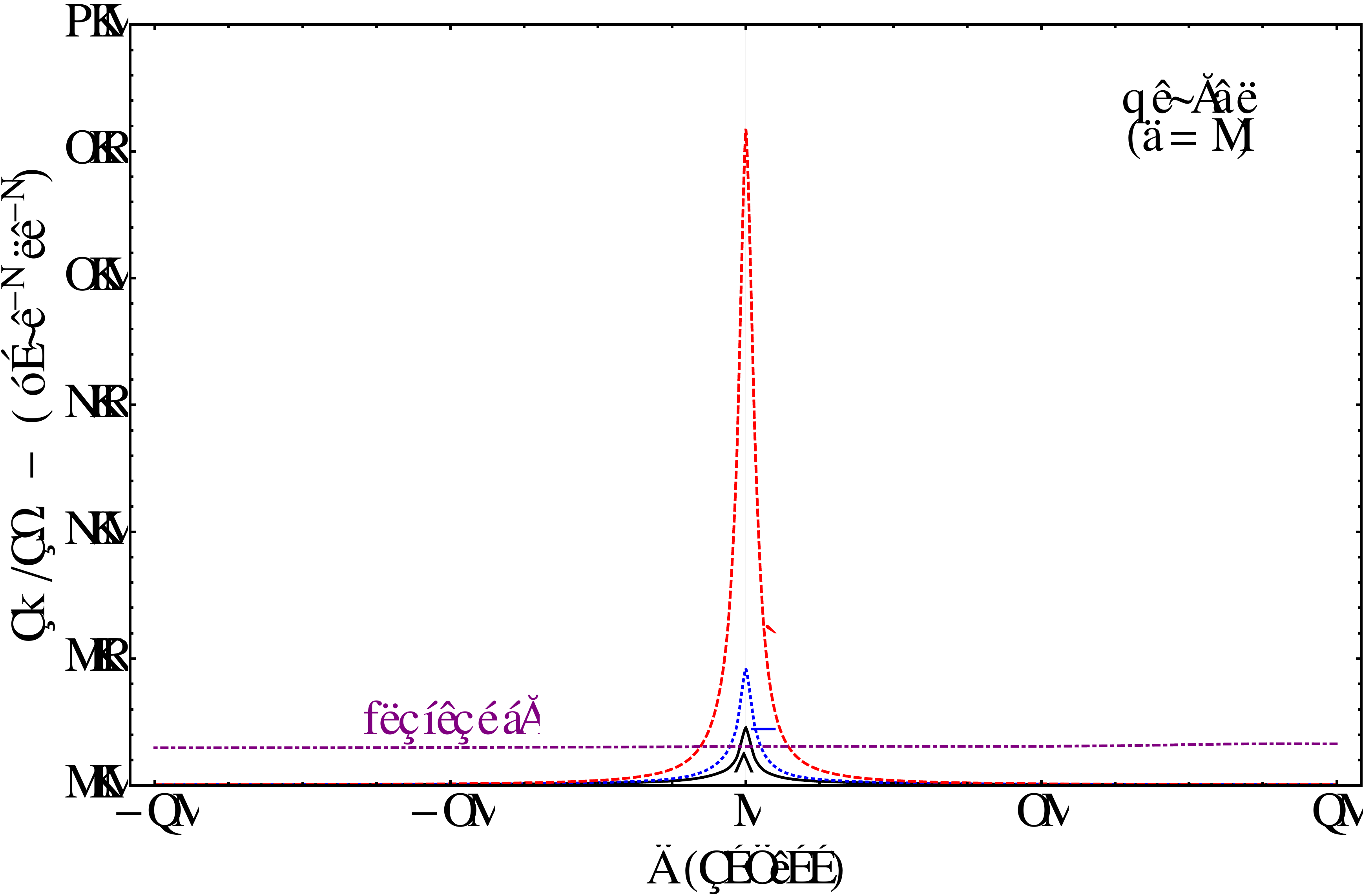} \\
\end{center}
\caption{ The HESE rate expected in IceCube as a function of the galactic longitude (left panels) 
and galactic latitude (right panels) for the three different models
considered in this work and for the isotropic flux that reproduces IceCube data. The legend of the different lines 
is that discussed in our previous plots. Upper panels refer to shower
events whereas lower panels refer to track events.   
\label{fig3}}
\end{figure*}

 The results of our calculations are shown in Fig.\ref{fig3}. 
We see that, due to the poor pointing accuracy, the showers produced 
by diffuse galactic neutrinos are diluted below the isotropic
component everywhere in the sky, except for the most favorable {\em Case C}.
On the contrary, the track rate remains dominant in a narrow 
region of the sky containing the galactic center. 
The expected track rate is, however, very small (see tab.~\ref{tab1}) 
making it difficult to obtain a non negligible detection probability.

We can estimate the chance of extracting the diffuse galactic
component from the HESE IceCube data sample, by evaluating
the fractional error $\delta N_{\rm I}$ in the determination of an 
excess $N_{\rm I}$  of track or shower events in a specific region 
of the sky with respect to the expectations $N_{\rm I,  iso}$ in the 
same region from an isotropic flux. 
We obtain: 
\begin{equation}
\delta N_{\rm I} \simeq \sqrt{\frac{1+\rho}{N_{\rm I}}}
\end{equation}
where ${\rm I=S,\, T}$, the paremeter $\rho= N_{\rm I, iso}/N_{\rm I}$
represents the background-to-signal ratio in the adopted observation window
and we neglected systematical error sources.
We consider for definiteness {\em Case C} since this is the only 
scenario in which we obtain a non negligible chance of detection 
due to the fact that it predicts a larger and more pronounced 
emission from the inner galactic region.
In this specific case, the optimal observation window for showers
is given by $ | l | \le 60^\circ $ and  $  |b| \le 15^\circ$, for which we obtain:
\begin{equation}
\delta N_{\rm S} = \frac{1.9}{\sqrt{T/1\,{\rm y}}}.
\end{equation}
For tracks, the optimal region is given by $ | l | \le
80^\circ $ and  $  |b| \le 3^\circ$ for which we have:
\begin{equation}
\delta N_{\rm T} = \frac{3.3}{\sqrt{T/1\,{\rm y}}}.
\end{equation}

The above results show that an observation time $T\ge 4 \,{\rm y}$ 
for showers and $T\ge 11\,{\rm y}$ for tracks is necessary 
to obtain $1\sigma$ hints for a galactic neutrino component, i.e. 
$\delta N_{\rm S}\le 1$ and/or $\delta N_{\rm T}\le 1$. 
For comparison, observation times larger than 35 years and 20 years 
are required to obtain a comparable significance for {\em Case A} and 
{\em Case B}, respectively.
This allow us to conclude that the detection of 
a statistically significant excess of events from the
galactic plane in present (or next future) IceCube
HESE data, as e.g. suggested by \cite{Neronov,Neronov2}, 
would require relatively large galactic fluxes, favoring scenarios 
similar to our {\em Case C} in which the CR density 
in the inner galactic region is greatly enhanced with respect to its
local value (see fig.\ref{fig1}).

\section{Conclusions}
\label{Conclusions}

In this paper, we have calculated the angular dependence of the
diffuse galactic neutrino flux and the corresponding IceCube HESE rate
by considering different assumptions for the CR density in the Galaxy. 
Namely, we have assumed that CR distribution is
homogenous in the Galaxy ({\em Case A}), that it follows the 
distribution of galactic CR sources ({\em Case  B}) and that 
it has a spectral index that depends on the galactocentric distance 
({\em Case C}).
Our conclusions are summarised in the following:
\vspace{0.3cm}

\noindent
{\em i)} In the considered scenarios, the angle-integrated galactic 
neutrino flux at 100 TeV  
is always subdominant with respect to the isotropic contribution
required to fit IceCube HESE data. However, it always
exists a region of the sky, that contains the galactic center, where
the galactic component is comparable or larger than the isotropic
contribution. \\
\noindent
{\em ii)} While the angle-integrated flux vary at most by a factor
$\sim 2$, the angular distribution of the diffuse galactic component is strongly
dependent on the assumed CR distribution. In perspective, this provides an
handle to discriminate among different scenarios, in an ideal detector 
with sufficient statistics and  good pointing capability;\\
\noindent
{\em iii)} The poor angular resolution for shower events and 
the smallness of the expected rate limit the possibility to extract
the diffuse galactic neutrino contribution from the IceCube HESE
data. In our analysis, only {\em Case C} has a non negligible
chance of detection, due to the fact that it predicts a larger and
more pronounced emission from regions close to the galactic center.
In the optimal region of the sky 
given by $ | l | \le 60^\circ $ and $  |b| \le 15^\circ$, we expect 
$\sim 2.5$ HESE events in four years of data-taking
that could be observed with $\sim 1 \sigma$ significance
above the isotropic contribution.
Note that the future KM3NeT \cite{KM3net} should be in 
better position, having the possibility to observe the inner galactic 
region with a relatively large exposure by using up-going passing
muons;\\
\noindent
{\em iv)} If a statistically significant excess of events from the
galactic plane will be observed in present or next future IceCube
HESE data, as e.g. suggested by \cite{Palladino,Neronov,Grasso}, 
this would favour models similar to our {\em Case C},  
in which the CR density in the inner galactic region is much larger than its local value, thus
bringing relevant information on the CR radial distribution.\\
\vspace{0.2cm}
 
As a final remark, since the major obstacle for the detection of
diffuse galactic neutrinos in IceCube is the smallness of the expected 
event rates (at level of $\sim 1\, {\rm y^{-1}}$ at most),
it would be interesting to explore the possibility of increasing the
statistics (at the level of $\sim {\rm few} \, {\rm y^{-1}}$ at least) by lowering
the detection  threshold, as it was done e.g. in \cite{bassa}
and \cite{ultimo}.





\begin{thebibliography}{99}

\bibitem{Aartsen:2013jdh}
  M.~G.~Aartsen {\it et al.} [IceCube Collaboration],
  Science {\bf 342} (2013) 1242856

\bibitem{Aartsen:2014gkd}
  M.~G.~Aartsen {\it et al.} [IceCube Collaboration],
  Phys.\ Rev.\ Lett.\  {\bf 113} (2014) 101101

\bibitem{Aartsen:2015ivb}
  M.~G.~Aartsen {\it et al.} [IceCube Collaboration],
  Phys.\ Rev.\ Lett.\  {\bf 114} (2015) no.17,  171102

\bibitem{Aartsen:2015rwa}
  M.~G.~Aartsen {\it et al.} [IceCube Collaboration],
  Phys.\ Rev.\ Lett.\  {\bf 115} (2015) no.8,  081102
  
\bibitem{Aartsen:2015knd}
  M.~G.~Aartsen {\it et al.} [IceCube Collaboration],
  Astrophys.\ J.\  {\bf 809} (2015) no.1,  98

\bibitem{Botner}
O.Botner (IceCube collaboration), in talks presented at the IPA Symposium 2015 (2015) 

\bibitem{Kistler:2006h}
  M.~D.~Kistler and J.~F.~Beacom,
  Phys.\ Rev.\ D {\bf 74} (2006) 063007

\bibitem{burgio}
  B.~Link and F.~Burgio,
  Phys.\ Rev.\ Lett.\  {\bf 94} (2005) 181101
  
\bibitem{Murase}
  K.~Murase,
  arXiv:1511.01590 [astro-ph.HE].

\bibitem{Lacki}
  B.~C.~Lacki, T.~A.~Thompson, E.~Quataert, A.~Loeb and E.~Waxman,
  Astrophys.\ J.\  {\bf 734} (2011) 107

\bibitem{Aartsen:2014cva}
  M.~G.~Aartsen {\it et al.} [IceCube Collaboration],
  Astrophys.\ J.\  {\bf 796} (2014) no.2,  109

\bibitem{ultimo}
  M.~G.~Aartsen {\it et al.} [IceCube Collaboration],
  arXiv:1605.00163 [astro-ph.HE].

\bibitem{waxman}
  E.~Waxman and J.~N.~Bahcall,
  Phys.\ Rev.\ D {\bf 59} (1999) 023002

\bibitem{tamborra}
  I.~Tamborra, S.~Ando and K.~Murase,
  JCAP {\bf 1409} (2014) 043

\bibitem{murasenew}
  K.~Murase, D.~Guetta and M.~Ahlers,
  Phys.\ Rev.\ Lett.\  {\bf 116} (2016) no.7,  071101

\bibitem{Ahlers} 
  M.~Ahlers, Y.~Bai, V.~Barger and R.~Lu,
  Phys.\ Rev.\ D {\bf 93} (2016) no.1,  013009

\bibitem{Palladino}
  A.~Palladino and F.~Vissani,
  arXiv:1601.06678 [astro-ph.HE].
  
\bibitem{Neronov}
  A.~Neronov and D.~Semikoz,
  Phys.\ Rev.\ D {\bf 93} (2016) no.12,  123002

\bibitem{Neronov2}
  A.~Neronov and D.~V.~Semikoz,
  Astropart.\ Phys.\  {\bf 75} (2016) 60

\bibitem{Troitsky}
  S.~Troitsky,
  JETP Lett.\  {\bf 102} (2015) no.12,  785
   [Pisma Zh.\ Eksp.\ Teor.\ Fiz.\  {\bf 102} (2015) 899]
  
\bibitem{berezinskii}
V. S. Berezinskii  {\it et al.} (1990), Astrophysics of cosmic rays.

\bibitem{Acero}
  F.~Acero {\it et al.} [Fermi-LAT Collaboration],
  Astrophys.\ J.\ Suppl.\  {\bf 223} (2016) no.2,  26

\bibitem{Morlino}
  S.~Recchia, P.~Blasi and G.~Morlino,
  arXiv:1604.07682 [astro-ph.HE].

\bibitem{Grasso}
  D.~Gaggero, D.~Grasso, A.~Marinelli, A.~Urbano and M.~Valli,
  Astrophys.\ J.\  {\bf 815} (2015) no.2,  L25

\bibitem{Evoli}
   C.~Evoli {\it et al.}, Phys.Rev.Lett. {\bf 108} (2012) 211102

\bibitem{Gaggero}
  D.~Gaggero et al., Phys.Rev. {\bf D91} (2015) no.8, 083012

\bibitem{Yang}
  R.~z.~Yang, F.~Aharonian and C.~Evoli,
  arXiv:1602.04710 [astro-ph.HE].

\bibitem{nuflavor}
  A.~Palladino, G.~Pagliaroli, F.~L.~Villante and F.~Vissani,
  Phys.\ Rev.\ Lett.\  {\bf 114} (2015) no.17,  171101

\bibitem{kelner}
  S.~R.~Kelner and F.~A.~Aharonian,
  Phys.\ Rev.\ D {\bf 78} (2008) 034013
   Erratum: [Phys.\ Rev.\ D {\bf 82} (2010) 099901]
  
\bibitem{Galprop}
http://galprop.stanford.edu/

\bibitem{SM}
  I.~V.~Moskalenko, A.~W.~Strong, J.~F.~Ormes and M.~S.~Potgieter,
  ApJ {\bf 565} (2002) 280-296

\bibitem{CREAM}
  Y.~S.~Yoon {\it et al.},
  Astrophys.\ J.\  {\bf 728} (2011) 122

\bibitem{KASCADE}
  T.~Antoni {\it et al.} [KASCADE Collaboration],
  Astropart.\ Phys.\  {\bf 24} (2005) 1
  
\bibitem{KASCADEGrande}
W.~D.~Apel {\it et al.},
  Astropart.\ Phys.\  {\bf 77} (2016) 21.

\bibitem{SNRs}
  D.~A.~Green,
  Mon.\ Not.\ Roy.\ Astron.\ Soc.\  {\bf 454} (2015) no.2,  1517
  doi:10.1093/mnras/stv1885
  [arXiv:1508.02931 [astro-ph.HE]].

\bibitem{Pulsars}
  D.~R.~Lorimer {\it et al.},
  Mon.\ Not.\ Roy.\ Astron.\ Soc.\  {\bf 372} (2006) 777

\bibitem{Icecube4}
  M.~G.~Aartsen {\it et al.} [IceCube Collaboration],
  arXiv:1510.05223 [astro-ph.HE].

\bibitem{Gaisser}
  T.~K.~Gaisser, T.~Stanev and S.~Tilav,
  Front.\ Phys.\ China {\bf 8} (2013) 748
  
\bibitem{IceCubeAeff}
IceCubeAeff

\bibitem{mis}
M. G. Aartsen {\it et al.} (IceCube Collaboration), 
Phys. Rev. D91, 022001 (2015)

\bibitem{PRD}
  A.~Ianni, G.~Pagliaroli, A.~Strumia, F.~R.~Torres, F.~L.~Villante and F.~Vissani,
  Phys.\ Rev.\ D {\bf 80} (2009) 043007
  doi:10.1103/PhysRevD.80.043007
  [arXiv:0907.1891 [hep-ph]].

\bibitem{bassa}
  M.~G.~Aartsen {\it et al.} [IceCube Collaboration],
  Phys.\ Rev.\ D {\bf 91} (2015) no.2,  022001
  
\bibitem{KM3net}
  S.~Adrian-Martinez {\it et al.} [KM3Net Collaboration],
  arXiv:1601.07459 [astro-ph.IM].

\end{thebibliography}
\end{document}